\documentclass[useAMS,usenatbib]{mn2e}
\usepackage{color}
\usepackage{graphicx}
\usepackage{mathrsfs}
\usepackage{epsfig}
\usepackage{natbib}
\usepackage{color}
\usepackage{float}
\usepackage{amsmath}
\usepackage{times}
\usepackage{upgreek}
\usepackage[varg]{txfonts}
\usepackage{enumerate}
\usepackage{verbatim}
\usepackage{booktabs}
\usepackage{multirow}
\usepackage{amssymb}
\usepackage{placeins}
\usepackage{array,longtable}
\usepackage{appendix}

\title[RAR in ETGs]{SDSS-IV MaNGA: A Distinct Mass Distribution Explored in Slow-Rotating Early-type Galaxies}
\author[Rong et al.]{Yu Rong$^{1,2,3}$\thanks{E-mail: rongyuastrophysics@gmail.com}, Hongyu Li$^{3}$, Jie Wang$^{3}$, Liang Gao$^{3,4}$, Ran Li$^{3}$, Junqiang Ge$^{3}$, \and Yingjie Jing$^{3}$, Jun Pan$^{3}$, J. G. Fern\'andez-Trincado$^{5,6}$, Octavio Valenzuela$^{7}$, \and Erik Aquino Ort\'iz$^{7}$\\
$^{1}$Instituto de Astrof\'isica, Pontificia Universidad Cat\'olica de Chile, Av. Vicu\~na Mackenna 4860, Macul, Santiago, Chile\\
$^{2}$Chinese Academy of Sciences South America Center for Astronomy and China-Chile Joint Center for Astronomy, Camino El Observatorio 1515,\\ Las Condes, Santiago, Chile\\
$^{3}$National Astronomical Observatories, Chinese Academy of Sciences, 20A Datun Road, Chaoyang District, Beijing 100012, China\\
$^{4}$Institute of Computational Cosmology, Department of Physics, University of Durham, Science Laboratories, South Road, Durham DH1 3LE, UK\\
$^{5}$Departamento de Astronom\'\i a, Casilla 160-C, Universidad de Concepci\'on, Concepci\'on, Chile\\
$^{6}$Institut Utinam, CNRS UMR6213, Univ. Bourgogne Franche-Comt\'e, OSU THETA , Observatoire de Besan\c{c}on, BP 1615, 25010 Besan\c{c}on Cedex, France\\
$^{7}$Instituto de Astronom$\imath$a, Universidad Nacional Autonoma de Mexico, A.P. 70-264, 04510, Mexico, D.F., Mexico
}

\begin{document}
\maketitle

\begin{abstract}
	
	We study the radial acceleration relation (RAR) for early-type galaxies (ETGs) in the SDSS MaNGA MPL5 dataset. The complete ETG sample show a slightly offset RAR from the relation reported by McGaugh et al. (2016) at the low-acceleration end; we find that the deviation is due to the fact that the slow rotators show a systematically higher acceleration relation than the McGaugh's RAR, while the fast rotators show a consistent acceleration relation to McGaugh's RAR. There is a $1\sigma$ significant difference between the acceleration relations of the fast and slow rotators, suggesting that the acceleration relation correlates with the galactic spins, and that the slow rotators may have a different mass distribution compared with fast rotators and late-type galaxies. We suspect that the acceleration relation deviation of slow rotators may be attributed to more galaxy merger events, which would disrupt the original spins and correlated distributions of baryons and dark matter orbits in galaxies.

\end{abstract}
\section{Introduction}

McGaugh et al. (2016) reported a correlation between the radial acceleration traced by rotation curves and that predicted by the observed distribution of baryons in a sample of rotation-supported disk galaxies, which is usually referred to as ``radial acceleration relation'' (RAR). Later, Lelli et al. (2017) extended this relation to the early-type galaxies (ETGs) including ellipticals and lenticulars, and dwarf spheroidals (dSphs). Although the relation scatters are relatively large in ETGs and dSphs, the different populations of galaxies follow virtually the same RAR. In general, the RAR reveals that the mass distributions of baryons and dark matters are tightly correlated, and may be independent of galactic classification.

However, note that the RAR was primarily obtained from rotating galaxies, but lack verification for a statistically robust sample of dispersion-supported galaxies. In the sample of Lelli et al. (2017), 17 ETGs are rotating lenticulars or disky ellipticals, while only 8 X-ray ETGs are approximately classical pressure-supported ellipticals. Recently, Chae et al. (2017) found that the nearly-round, pure-bulge elliptical galaxies do not follow the RAR, by studying the radial accelerations of $\sim 7000$ Sloan Digital Sky Survey DR7 galaxies. Yet their sample should also include many face-on prolate or oblate rotation-supported ETGs. Later, \cite{Bilek17} investigated a small sample of 15 ETGs, and found that only the 4 fast rotators follow the RAR for the disk galaxies. However, their fast rotators are disky isophotes, appear very elongated, and might be spiral galaxies which lost their gas, suggesting that their results may not be a representative for all of the ETGs. Indeed, whether the RAR for the disk galaxies is also universal for ETGs is still under debate.

The Sloan Digital Sky Survey Mapping Nearby Galaxies at Apache Point Observatory (SDSS MaNGA, Bundy et al. 2015) is the largest integral-field spectroscopy (IFS) survey covering about 10~000 large, nearby galaxies with a median full width at half maximum (FHWM) of PSF of $\sim 2.5''$, and can provide well-resolved and high-quality stellar and gas kinematic maps and enable us to study the total mass distribution from dynamical modelling and stellar mass distribution using stellar population synthesis (SPS) predictions. The field-of-view of MaNGA covers a $1.5\--2.5$ effective radius range, which allows us to study the mass distribution from the galaxy cores to the halo regions. The current Fifth MaNGA Product Launch (MPL5; Abolfathi et al. 2017) released 2778 galaxies of both the early- and late-type galaxies, which permit us to precisely test the radial accelerations of baryons ($g_{\rm{bar}}$) and total masses ($g_{\rm{tot}}$) of the ETGs.

In this work, we study the RAR of ETGs. In section 2, we select the ETG sample from the MaNGA MPL5 catalog, and estimate their total and stellar masses from Jeans dynamical modelling and SPS, respectively. In section 3, we test the correlation between the radial accelerations of the total and baryonic masses of ETGs, and compare the acceleration relations of slow and fast rotators. We discuss the results in section 4. Throughout this letter, we use a $\Lambda$CDM cosmology with $\Omega_{\rm{M}}=0.272, \Omega_{\Lambda}=0.728$, and $H_0=70\ \rm km/s/Mpc$, use ``dex'' to mean the anti-logarithm, i.e. 0.1~dex=$10^{0.1}$=1.258, and use $\log$ to mean $\log_{10}$. 

\section{The Data}

\subsection{ETG sample selection}

The MaNGA MPL5 spectra are extracted by using the official data reduction pipeline (DRP, Law et al. 2016), and kinematical data are then extracted using the official data analysis pipeline (DAP, Westfall et al. in prep). See the references for more details about the MaNGA instrumentation (Drory et al. 2015), observing strategy (Law et al. 2015), spectrophotometric calibration (Yan et al. 2016a), and survey execution and initial data quality (Yan et al. 2016b). In order to obtain the stellar kinematics, the data cubes are spatially Voronoi binned (Cappellari \& Copin 2003) to a continuum signal-to-noise ratio S/N=10. In each spaxel, the stellar line-of-sight velocity $V_{\rm{s}}$ and dispersion $\sigma_{\rm{s}}$ are calculated from the spectrum. For each galaxy, the kinematic major axis of the stellar component is then obtained by using the {\bf{\textsc{fit\_kinematic\_pa}}} software \citep{Krajnovic06}.

We search the NASA-Sloan Atlas catalog{\footnote{http://www.sdss.org/dr13/manga/manga-target-selection/nsa/}} (NSA v1\_0\_1, Blanton et al. 2011) to obtain the galactic K-corrected rest-frame magnitudes, colors (e.g., NUV-$r$, $g-r$), $r$-band S\'ersic indices $n$, and stellar masses $M_{\star}$. 
The ellipticities $\epsilon$ are obtained by using the {\bf{\textsc{find\_galaxy}}} software \citep{Cappellari06} to fit their $r$-band images.

Since we are interested in whether the acceleration relation changes with galactic spins, we calculate the dimensionless spin parameter $\lambda$ defined in \cite{Emsellem07} to distinguish fast and slow rotators, i.e.,
\begin{equation} 	\lambda=\frac{\Sigma_{i=1}^{N_p}F_iR_i|V_{{\rm{s}},i}|}{\Sigma_{i=1}^{N_p}F_iR_i\sqrt{V_{{\rm{s}},i}^2+\sigma_{{\rm{s}},i}^2}},
\label{spin}
\end{equation} 
where $F_i$ is the flux inside the $i$th bin, $R_i$ is the distance of the $i$th bin to the galaxy center, and $V_{{\rm{s}},i}$ and $\sigma_{{\rm{s}},i}$ are the mean stellar velocity and dispersion in the $i$th bin, respectively. $\lambda/\sqrt\epsilon$ can assess the galaxy rotation, e.g., $\lambda/\sqrt\epsilon>0.31$ corresponds to a fast rotator (Emsellem et al. 2011).

Their stellar mass distributions will be inferred from SPS models with a \citet{Salpeter55} initial mass function (IMF), and their total mass distributions will be obtained from Jeans anisotropic modelling (JAM; Cappellari 2008). The gas mass can be evaluated from the relation between the cold gas-to-stellar mass ratio (G/S) and color, as well as axial ratio \citep{Eckert15}, i.e. log~G/S=$-1.002\times(3.563(g-r)+0.534(b/a))+1.813$. This relation is robust for the galaxies with $3.563(g-r)+0.534(b/a)<2.6$, and G/S is negligible for the galaxies with $3.563(g-r)+0.534(b/a)>2.6$.

ETGs are selected with the following criteria:

1) we remove a galaxy if there are less than 300 Voronoi bins with S/N$\simeq 10$;

2) the mergers, irregulars, and strong barred galaxies are removed;

3) the kinematic center of a galaxy coincides with the centroid of the stellar distribution;

4) morphology: the S\'ersic index of each ETG is $n>2.5$;

5) ellipticity: since the total mass from dynamic modelling can not be accurately recovered if $\epsilon\sim 0$ (Lablanche et al. 2012), therefore, we remove the galaxies with $\epsilon<0.2$; 

6) color: a typical ETG should be red, passive, and lack of star formation (i.e. lack of gas). The near-ultraviolet band (NUV) is a better indicator for star formation, compared with the optical bands; we only select the galaxies with colors of NUV-$r>3.7$;

7) G/S: we remove the gas-rich galaxies with $3.563(g-r)+0.534(b/a)<2.6$.

\begin{figure*}
\centering
\includegraphics[scale=0.48]{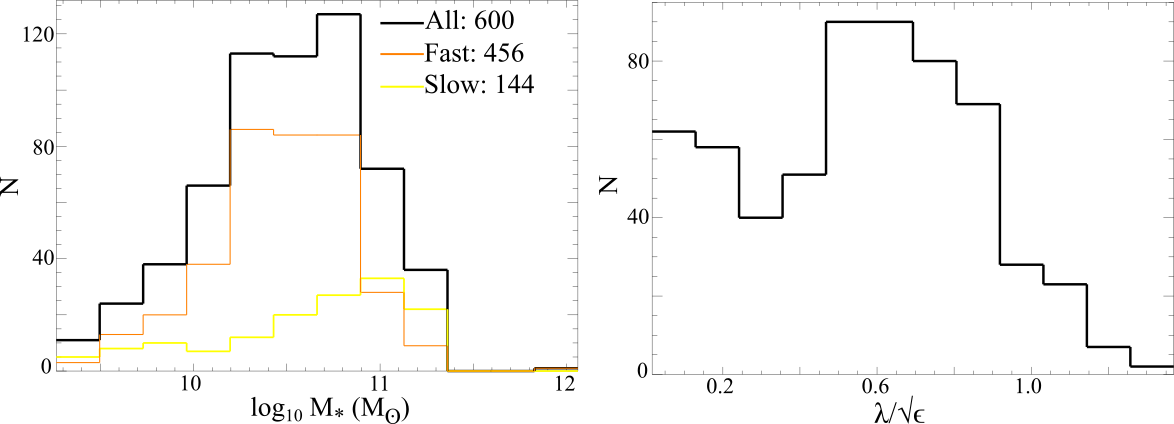}
\caption{The distributions of $M_{\star}$ (left panel) and $\lambda/\sqrt\epsilon$ (right panel) for the complete ETG sample (black), slow rotators ($\lambda/\sqrt{\epsilon}<0.31$, yellow), and fast rotators ($\lambda/\sqrt{\epsilon}>0.31$, orange)}
\label{stellarmass_spin}
\end{figure*}

Finally, 600 ETGs are selected, and the distributions of their $M_{\star}$ and $\lambda/\sqrt\epsilon$ are explored in Fig.~\ref{stellarmass_spin}. 

\subsection{$g_{\rm{tot}}$ from the dynamical modelling}
\label{sec:JAM}
We perform JAM for all the sample galaxies. The modelling allows for anisotropy in the second-velocity-moments, and provides accurate ($\sim 10\%-18\%$) and unbiased estimates of the total mass distribution as tested in \citet{Li16}.
We first use the Multi-Gaussian Expansion (MGE) method \citep{Emsellem94} with the fitting algorithm and Python 
software{\footnote{Available from http://purl.org/cappellari/software}} by \cite{Cappellari02} to fit the SDSS $r$-band image, and then deproject the surface brightness to obtain the luminosity density by assuming an inclination.
The deprojected stellar luminosity density is used as the tracer density in the modelling.
Similar to the equation~(2) in \citet{Poci17}, we assume that the total mass density is axisymmetric and follows
\begin{equation}
\label{eq:gnfw}
        \rho_{\rm tot}(l)=\rho_s \left(\frac{l}{l_s}\right)^\gamma
            \left(\frac{1}{2}+\frac{1}{2}\frac{l}{l_s}\right)^{-\gamma-3},
\end{equation}
where $z$-axis and $R$ denote the symmetric axis and radius in the $z$-plane, respectively, and $l=\sqrt{R^2+z^2/q^2}$ is the elliptical radius; $q$, $l_s$, and $\rho_s$ are the intrinsic axis ratio, scale radius, and density at the scale radius, respectively; $\gamma$ is the inner slope.

The model has six free parameters: (i) inclination, (ii) velocity anisotropy, $\beta_z$, (iii) $l_s$, (iv) $\rho_s$, (v) $\gamma$, and (vi) $q$. Running JAM within a Markov-Chain-Monte-Carlo framework
({\bf{\textsc{{emcee}}}}{\footnote{http://dfm.io/emcee/current/}}, \citealt{Foreman-Mackey2013}), we obtain the best-fitting parameters which give the best model matching the observed second-velocity-moment map. The $g_{\rm{tot}}(R)$ on the equatorial plane (i.e. $z=0$) is then calculated from the JAM estimated total mass distribution. 

\subsection{$g_{\rm{bar}}$ from stellar population synthesis}

For each galaxy, $g_{\rm{bar}}(R)$ is calculated from the 3-dimensional (3D) stellar mass distribution constructed by deprojecting the 2-dimensional (2D) stellar mass surface density which is calculated from the 2D light distribution from the MGE and stellar mass-to-light ratio $M_{\star}/L$ map from SPS models, using the inclination derived from the best-fitting JAM parameters. We first use the {\bf{\textsc{{\scriptsize p}pxf}}} \citep{Cappellari04,Cappellari17} with the MILES-based (S\'anchez-Bl\'azquez et al. 2006) SPS models of Vazdekis et al. (2010), to evaluate $M_{\star}/L$ in each bin. The \citet{Salpeter55} IMF and \citet{Calzetti00} extinction laws are assumed. Before spectrum fitting, the data cubes are Voronoi binned \citep{Cappellari03} to S/N=30. Following \citet{Li17a}, the stellar mass-to-light ratio in the $i$-th bin is calculated as
\begin{equation}
(M_{\star}/L)_{i}=(\Sigma_{j=1}^N w_jM_j)/(\Sigma_{j=1}^N w_jL_j), 
\end{equation}
where $M_j$, $L_j$, and $w_j$ are the stellar mass, $r$-band extinction-corrected luminosity, and weight of the $j$-th template from {\bf{\textsc{{\scriptsize p}pxf}}} fitting \citep{Cappellari13}, respectively. After obtaining the $M_{\star}/L$ map, we use the same manner described in section 2.4 of Poci et al. (2017) to construct the 3D stellar mass distribution. The $g_{\rm{bar}}(R)$ on the equatorial plane is calculated from the stellar mass distribution.

\section{Radial Acceleration Relation}

\begin{figure*}
\centering
\includegraphics[scale=0.45]{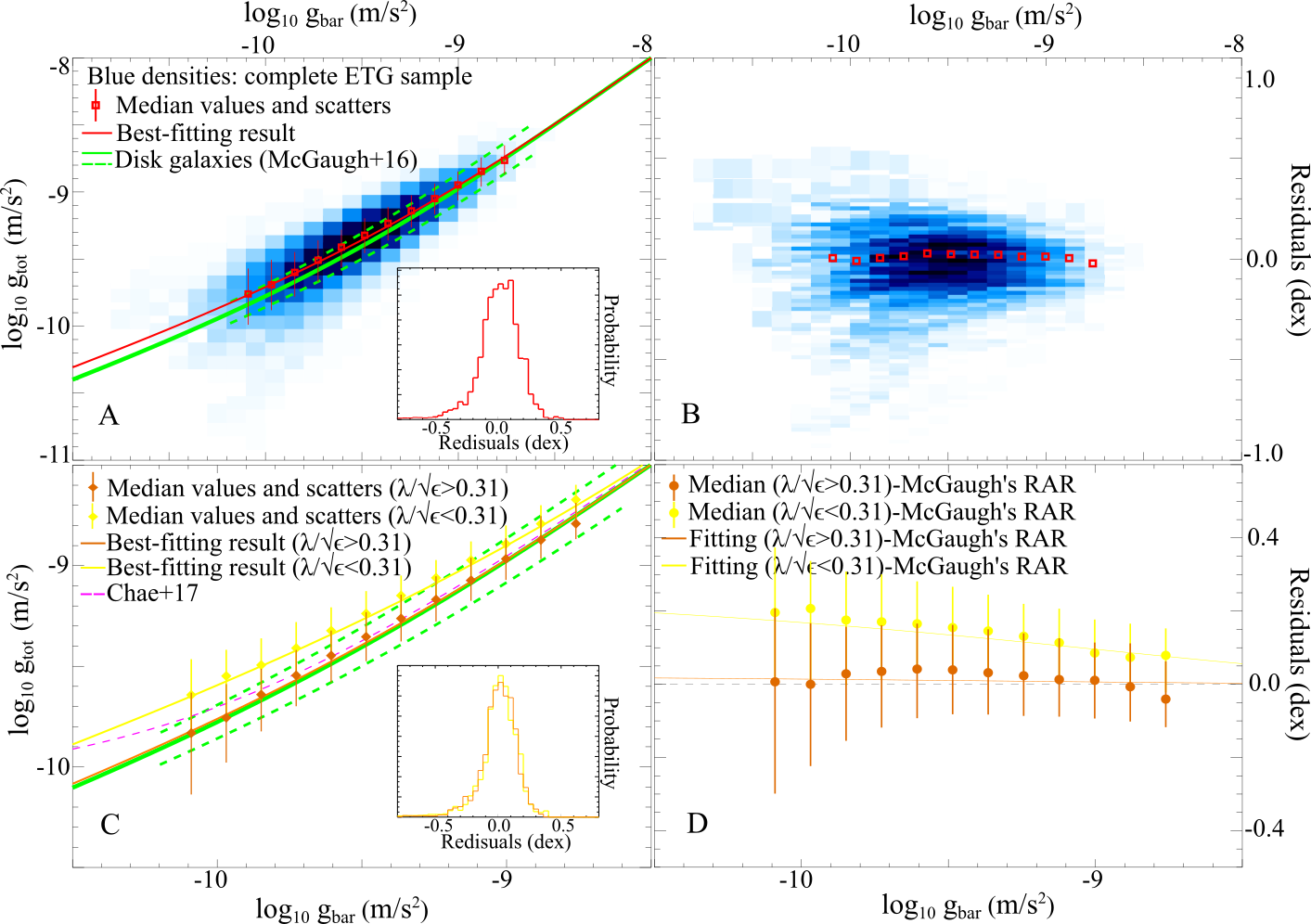}
\caption{Panel A: $g_{\rm{tot}}\--g_{\rm{bar}}$ probability distribution of the complete ETG sample is shown by the blue density distribution. The green curves denote the RAR and its $1\sigma$ scatter reported by McGaugh et al. (2016). The red squares and corresponding error bars show the median values of $g_{\rm{tot}}$ in the $\log g_{\rm{bar}}$ bins, and the red curve explores the fitting result with Eq.~(\ref{ggfitting}). The inset shows the distribution of residuals around the best-fitting relation. Panel B: the residuals as a function of $g_{\rm{bar}}$, and the blues density distribution shows the probability distribution of the residuals. The red squares show the median values of the residuals in the $\log g_{\rm{bar}}$ bins. Panel C: analogous to panel A, it shows the results of the slow (yellow) and fast (orange) rotators, respectively. The magenta dashed curve presents the fitting result of Chae et al. (2017) for the nearly-round, pure-bulge elliptical galaxies. Panel D: the colored circles (the error bars show $1\sigma$ scatters) and curves show the median values and best-fitting results in panel C substracting the RAR of McGaugh et al. (2016), respectively. The black dashed line denotes residuals equal to 0.}
\label{RAR}
\end{figure*}

For each galaxy, we exclude the accelerations in the innermost region ($R<3$~arcsec), since the enclosed mass distribution cannot be accurately contructed due to the limited spatial resolution. We then uniformly generate $20\--60$ sets (depending on the observed radius range; on average about 40 sets in one galaxy) of ($g_{\rm{tot}}$, $g_{\rm{bar}}$) from $R=3$~arcsec to the observed maximum radius of a galaxy. The $g_{\rm{tot}}$ and $g_{\rm{bar}}$ for the complete sample of ETGs are shown by the blue density distribution (a bluer shading indicates a larger number of points in this region) in panel A of Fig.~\ref{RAR}. The median $g_{\rm{tot}}$ and $1\sigma$ scatter in each $\log g_{\rm{bar}}$ bin are shown by a red square and corresponding error bar in panel A, respectively. We then fit the points of the complete ETG sample with the equation,
\begin{equation}
	g_{\rm{tot}}=\frac{g_{\rm{bar}}}{1-\exp{(-\sqrt{g_{\rm{bar}}/g_{\dagger}})}},
\label{ggfitting}
\end{equation}
where $g_{\dagger}$ is the fitting parameter \citep{McGaugh16}. The best-fitting result, as explored by the red curve in panel A, well coincides with the median $g_{\rm{tot}}$. The fitting residuals and their distribution are also shown as the blue density distributions (the red squares denote the median residuals) in panel B and inset in panel A, respectively. We find that the fitting result is robust.

For comparison, the RAR obtained from the rotation-supported disk galaxies and its $1\sigma$ scatter, as reported by \cite{McGaugh16}, are also highlighted by the green solid and dashed components, respectively. We find that the best-fitting result is slightly higher than McGaugh's RAR for disk galaxies at the low-acceleration end. In order to test whether the difference is due to the dispersion-dominated ETGs, in panel C of Fig.~\ref{RAR}, we plot the acceleration relations for the slow rotators with $\lambda/\sqrt\epsilon < 0.31$ (yellow) and fast rotators with $\lambda/\sqrt\epsilon > 0.31$ (orange), respectively. The slow and fast rotators occupy about $24\%$ and $76\%$ of the complete sample, respectively. Their median $g_{\rm{tot}}$ (colored circles) and $1\sigma$ scatters (colored error bars) in the $\log g_{\rm{bar}}$ bins, and best-fitting results (colored curves) with Eq.~(\ref{ggfitting}) are also shown by the corresponding colors in panel C. In panel D, we plot the differences of the median $g_{\rm{tot}}$ (colored circles) and fitting results (colored solid curves) from McGaugh's RAR, for the two samples.

We find that there is a modest deviation between the acceleration relation of the slow rotators and McGaugh's RAR. The acceleration relation of the slow rotators is offset towards higher $g_{\rm{tot}}$ values relative to McGaugh's RAR at a given $g_{\rm{bar}}$; at the low-acceleration end $\log g_{\rm{bar}}\simeq -10.0$, the median $g_{\rm{tot}}$ of the slow rotators is about 0.2~dex higher (with a 1 $\sigma$ scatter of $\sim 0.2$~dex) than McGaugh's RAR, while the deviation tends to decrease with increasing $g_{\rm{bar}}$. Yet the acceleration relation of the fast rotators is approximately consistent with McGaugh's RAR. There is a difference between the acceleration relations of the slow and fast rotators at about $1\sigma$ significance in the range of $-10.0< \log g_{\rm{bar}}<-9.0$, e.g., at $\log g_{\rm{bar}}\sim -10.1$, the median $\log g_{\rm{tot}}$ of the slow rotators is about $-9.7_{-0.1}^{+0.2}$, yet the fast rotators have $\log g_{\rm{tot}}\sim -9.9_{-0.3}^{+0.2}$; at $\log g_{\rm{bar}}\sim -9.2$, the median $\log g_{\rm{tot}}$ of the slow and fast rotators are $-9.1_{-0.1}^{+0.1}$ and $-9.2_{-0.1}+{0.1}$, respectively; therefore, the median $\log g_{\rm{tot}}$ of the slow rotators lie outside of the 1 $\sigma$ scatter from the median $\log g_{\rm{tot}}$ of the fast rotators, and vice versa. The modest RAR deviation of the slow rotators suggests a correlation between the acceleration relations and galactic spins, and possibly implies that the slow rotators may have a different mass distribution compared with fast rotators, dSphs, and LTGs.

In panel C, we also plot the log-linear fitting result (i.e. $\log[(g_{\rm{tot}}-g_{\rm{bar}})/g_{\rm{bar}}]=-0.93\log[g_{\rm{bar}}/g_0]-0.08$, where $g_0=1.2\times 10^{-10}\ \rm{m/s^2}$; shown as the magenta component) obtained from the nearly-round, pure-bulge elliptical galaxies by \cite{Chae17}. The acceleration relation of the slow rotators in this work deviates from McGaugh's RAR more significantly than the finding in \cite{Chae17}, which may be due to the fact that the sample of \cite{Chae17} includes many nearly-face-on fast rotators. 


\section{Discussion}

\begin{figure*}
\centering
\includegraphics[scale=0.8]{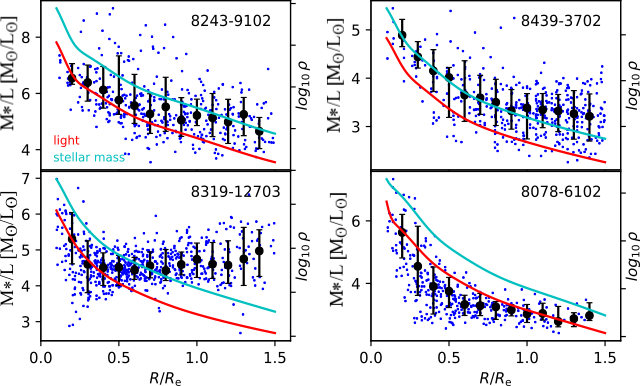}
\caption{The top and bottom panels show the examples of the fast and slow rotators, respectively. In each panel, the blue points show the stellar-to-light ratios of the different Voronoi bins of a galaxy from SPS (in the $r$-band); the black points and the error bars show the median stellar-to-light ratio and 1 $\sigma$ scatter in each radius bin. The trends of the 3D light and stellar mass profiles are highlighted by the red and cyan components, respectively; the light and mass densities show the mean values at $R/R_{\rm{e}}$, and are in a unit of $L_{\odot}/{\rm{pc}}^3$ and $M_{\odot}/{\rm{pc}}^3$, respectively.}
\label{ML_radius}
\end{figure*}

In our work, $g_{\rm{tot}}$ is obtained from JAM, in the sense that an ETG is assumed in equilibrium. Note that a typical ETG is gas-poor, and the stars are collisionless and thus less susceptible to perturbations, therefore, the assumption of equilibrium should be reasonable for the ETGs. $g_{\rm{bar}}$ depends on the assumed $M_{\star}/L$, and thus is susceptible to the IMFs, stellar ages, and metallicities. For the ETGs with large velocity dispersions, it is better to assume a Salpeter IMF rather than a Kroupa (2001) or Chabrier (2003) IMF \citep[e.g.,][]{Thomas11,Cappellari12,Dutton12,Li17a}. Here we show the pixel-to-pixel stellar-to-light ratios of several ETG examples. As shown in Fig.~\ref{ML_radius}, the blue points show the stellar-to-light ratios, $(M_*/L)_i$, from SPS in the different Voronoi bins (in the $r$-band). We also calculate the median $M_*/L(R)$ in each radius bin (shown as the black points), which is used to estimate the distribution of the stellar mass of an ETG in this work. Note that the trends of our radially variable stellar-to-light ratios are similar to the results of ETGs shown by van Dokkum et al. (2017). The 3D light and stellar mass profiles are also highlighted in Fig.~\ref{ML_radius}.

The acceleration relation scatter for the ETGs at the low-acceleration end is larger than the scatter of the rotation-supported disk galaxies, as explored by McGaugh et al. (2016), yet all relations are in agreement within the large scatter of ETGs reported by Lelli et al. (2017). We note that the large scatters of ETGs may be partly attributed to the intrinsic deviations of the acceleration relations between the fast and slow rotators.

\begin{figure}
\centering
\includegraphics[scale=0.22]{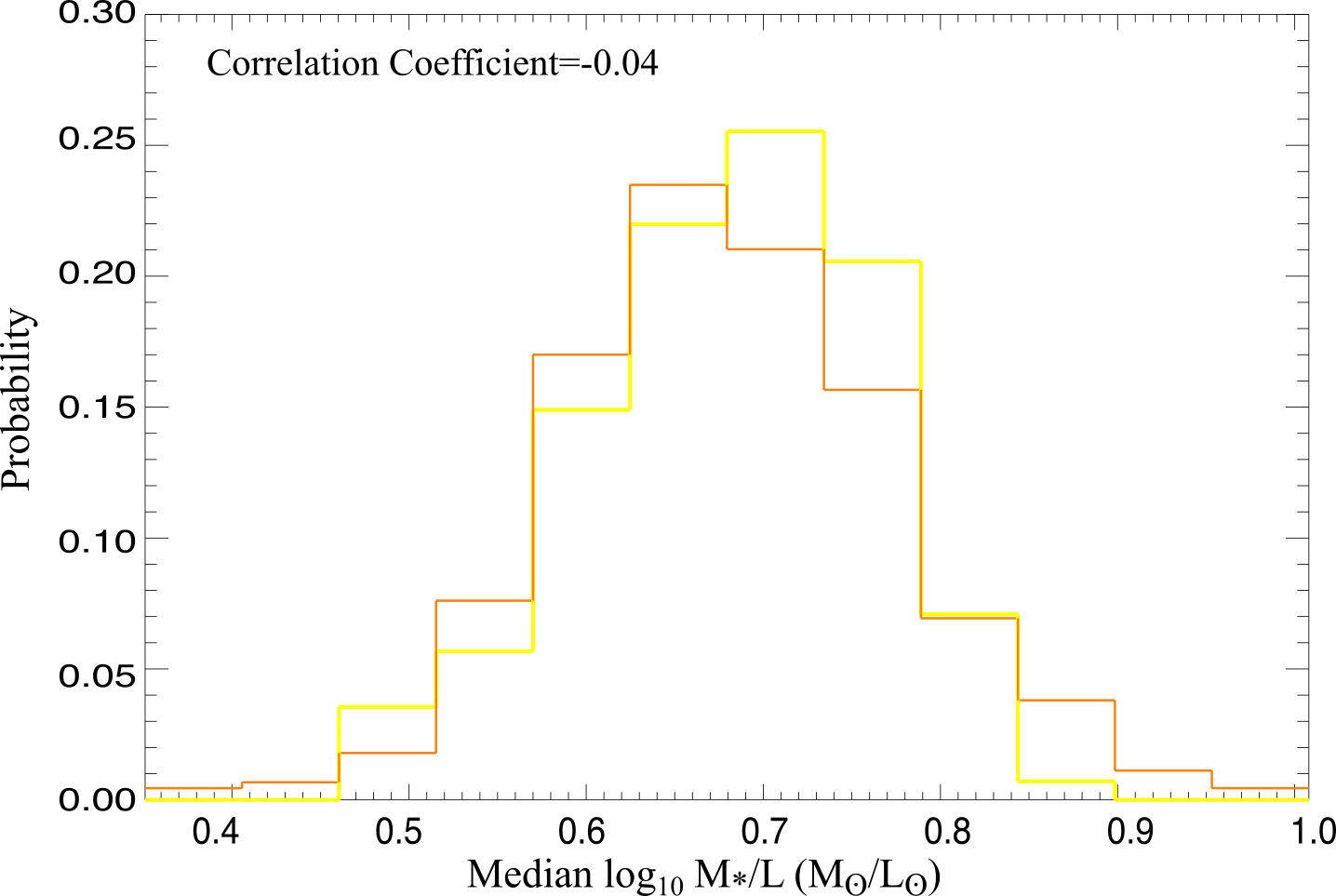}
\caption{The distributions of the median stellar-to-light ratios within the effective radii of the slow (yellow) and fast (orange) rotators. The correlation coefficient between $M_*/L$ and $\lambda/\sqrt{\epsilon}$ is close to 0.}
\label{ML_correlation}
\end{figure}

The physical mechanism leading to the $1\sigma$ difference between the acceleration relations of the slow and fast rotators is unclear. We plot the distributions of the median $M_{\star}/L$ in the effective radii of the slow and fast rotators in Fig.~\ref{ML_correlation}, and find that their distributions have no significant difference. We also find that there is no correlation between the median $M_{\star}/L$ and $\lambda/\sqrt\epsilon$ (correlation coefficient is close to 0), in the sense that the RAR deviation between the fast and slow rotators should not be attributed to the possible difference between the $M_*/L$ of the two samples. We also find that the stellar masses of the slow rotators are marginally higher than those of the fast rotators (as shown in the left panel of Fig.~\ref{stellarmass_spin}), which may suggest that the slow rotators have undergone more galaxy merger events \citep{Cappellari16}. Indeed, galaxy mergers can disrupt galaxy spins \citep{Penoyre17,Li17b} and original coupled distributions between baryons and dark matter, plausibly leading to the deviation of the acceleration relation of the slow rotators. For instance, during a gas-rich major merger, the stars and dark matter of the binary show no or weak viscosities and quickly become equilibrium, yet the gas shows strong viscosity and thus needs a period of time to be relaxed, leading to a higher $g_{\rm{bar}}$ in the relatively outer regions of halos, i.e. the low-$g_{\rm{tot}}$ end. Another possible way is to kick stars from the central regions by binary supermassive black holes in dissipationless major mergers, which also lead to mass deficits in cores of slow rotators \citep{Kormendy09}.

\section*{Acknowledgments}

YR thanks Thomas H. Puzia, Dandan Xu, Zheng Zheng, Yougang Wang, Yan-Mei Chen, Xiao-Lin Yu, Zhou Fan, Jin-Yuan Liao, Shuangpeng Sun for their help and discussions, and acknowledges supports from CAS-CONICYT postdoctoral fellowship No. 16004 and NSFC grant No. 11703037. LG acknowledges support from NSFC grants (nos. 11133003 and 11425312). THP acknowledges support in form of the FONDECYT Regular Project No. 1161817 and by the BASAL Center for Astrophysics and Associated Technologies (PFB-06). JP acknowledges support from the National Basic Research Program of China (program 973 under grant No. 2015CB857001) and NSFC under grant No. 11573030. This research made use of Marvin, a core Python package and web framework{\footnote{https://sas.sdss.org/marvin2/}} for MaNGA data, developed by Brian Cherinka, Jos\'e S\'anchez-Gallego, and Brett Andrews. (MaNGA Collaboration, 2017). Funding for the Sloan Digital Sky Survey IV has been provided by the Alfred P. Sloan Foundation, the U.S. Department of Energy Office of Science, and the Participating Institutions. SDSS acknowledges support and resources from the Center for High-Performance Computing at the University of Utah. The SDSS web site is www.sdss.org.

SDSS is managed by the Astrophysical Research Consortium for the Participating Institutions of the SDSS Collaboration including the Brazilian Participation Group, the Carnegie Institution for Science, Carnegie Mellon University, the Chilean Participation Group, the French Participation Group, Harvard-Smithsonian Center for Astrophysics, Instituto de Astrof\'isica de Canarias, The Johns Hopkins University, Kavli Institute for the Physics and Mathematics of the Universe (IPMU) / University of Tokyo, Lawrence Berkeley National Laboratory, Leibniz Institut für Astrophysik Potsdam (AIP), Max-Planck-Institut f\"ur Astronomie (MPIA Heidelberg), Max-Planck-Institut f\"ur Astrophysik (MPA Garching), Max-Planck-Institut f\"ur Extraterrestrische Physik (MPE), National Astronomical Observatories of China, New Mexico State University, New York University, University of Notre Dame, Observat\'orio Nacional / MCTI, The Ohio State University, Pennsylvania State University, Shanghai Astronomical Observatory, United Kingdom Participation Group, Universidad Nacional Aut\'onoma de M\'eXico, University of Arizona, University of Colorado Boulder, University of Oxford, University of Portsmouth, University of Utah, University of Virginia, University of Washington, University of Wisconsin, Vanderbilt University, and Yale University.

\bibliographystyle{mn2e}

\begin{thebibliography}{}

\bibitem[\protect\citeauthoryear{Abolfathi et al.}{2017}]{Abolfathi17} Abolfathi B., et al., 2017, arXiv: 1707.09322
\bibitem[\protect\citeauthoryear{B\'ilek \& Samurovi\'c}{2017}]{Bilek17} B\'ilek M., Samurovi\'c S., 2017, arXiv: 1711.06335
\bibitem[\protect\citeauthoryear{Blanton et al.}{2011}]{Blanton11} Blanton M. R., Kazin E., Muna D., Weaver B. A., Price-Whelan A., 2011, AJ, 142, 31
\bibitem[\protect\citeauthoryear{Bundy et al.}{2015}]{Bundy15} Bundy K., et al., 2015, ApJ, 798, 7
\bibitem[\protect\citeauthoryear{Buote et al.}{2016}]{Buote16} Buote D. A., Su Y., Gastaldello F., Brighenti F., 2016, ApJ, 826, 146
\bibitem[\protect\citeauthoryear{Calzetti}{2000}]{Calzetti00} Calzetti D., Armus L., Bohlin R. C., Kinney A. L., Koornneef J., Storchi-Bergmann T., 2000, ApJ, 533, 682
\bibitem[\protect\citeauthoryear{Cappellari}{2002}]{Cappellari02} Cappellari M., 2002, MNRAS, 333, 400 
\bibitem[\protect\citeauthoryear{Cappellari}{2008}]{Cappellari08} Cappellari M., 2008, MNRAS, 390, 71
\bibitem[\protect\citeauthoryear{Cappellari}{2016}]{Cappellari16} Cappellari M., 2016, ARA\&A, 54, 597
\bibitem[\protect\citeauthoryear{Cappellari \& Copin}{2003}]{Cappellari03} Cappellari M., Copin Y., 2003, MNRAS, 342, 345
\bibitem[\protect\citeauthoryear{Cappellari \& Emsellem}{2004}]{Cappellari04} Cappellari M., Emsellem E., 2004, PASP, 116, 138
\bibitem[\protect\citeauthoryear{Cappellari et al.}{2006}]{Cappellari06} Cappellari M., et al., 2006, MNRAS, 366, 1126
\bibitem[\protect\citeauthoryear{Cappellari et al.}{2012}]{Cappellari12} Cappellari M., et al., 2012, Nature, 484, 485
\bibitem[\protect\citeauthoryear{Cappellari et al.}{2013}]{Cappellari13} Cappellari M., et al., 2013, MNRAS, 432, 1862
\bibitem[\protect\citeauthoryear{Cappellari et al.}{2017}]{Cappellari17} Cappellari M., 2017, MNRAS, 466, 798
\bibitem[\protect\citeauthoryear{Catinella et al.}{2013}]{Catinella13} Catinella B., et al., 2013, MNRAS, 436, 34
\bibitem[\protect\citeauthoryear{Catinella et al.}{2010}]{Catinella10} Catinella B., Schiminovich D., Kauffmann G., et al., 2010, MNRAS, 403, 683
\bibitem[\protect\citeauthoryear{Chabrier}{2003}]{Chabrier03} Chabrier G., 2003, ApJ, 586, L133
\bibitem[\protect\citeauthoryear{Chae et al.}{2017}]{Chae17} Chae K., Bernardi M., Sheth R. K., 2017, arXiv: 1707.08280
\bibitem[\protect\citeauthoryear{Dai et al.}{2010}]{Dai10} Dai X., Bregman J. N., Kochanek C. S., Rasia E., 2010, ApJ, 719, 119
\bibitem[\protect\citeauthoryear{Drory et al.}{2015}]{Drory15} Drory N., et al., 2015, AJ, 149, 77
\bibitem[\protect\citeauthoryear{Dutton et al.}{2012}]{Dutton12} Dutton A. A., Mendel J. T., Simard L., 2012, MNRAS, 422L, 33
\bibitem[\protect\citeauthoryear{Eckert et al.}{2015}]{Eckert15} Eckert K. D., Kannappan S. J. Stark D. V., Moffett A. J., Norris M. A., Snyder E. M., Hoversten E. A., 2015, ApJ, 810, 166
\bibitem[\protect\citeauthoryear{Emsellem et al.}{1994}]{Emsellem94} Emsellem E., Monnet G., Bacon R., 1994, A\&A, 285, 723
\bibitem[\protect\citeauthoryear{Emsellem et al.}{2007}]{Emsellem07} Emsellem E., et al., 2007, MNRAS, 379, 401
\bibitem[\protect\citeauthoryear{Foreman-Mackey et al.}{2013}]{Foreman-Mackey2013} Foreman-Mackey D., Hogg D.~W., Lang D., Goodman J., 2013, PASP, 125, 306 
\bibitem[\protect\citeauthoryear{Gonzalez et al.}{2013}]{Gonzalez13} Gonzalez A. H., Sivanandam S., Zabludoff A. I., Zaritsky D., 2013, ApJ, 778, 14
\bibitem[\protect\citeauthoryear{Graham \& Driver}{2005}]{Graham05} Graham A. W., Driver S. P., 2005, PASA, 22, 118
\bibitem[\protect\citeauthoryear{Humphrey et al.}{2012}]{Humphrey12} Humphrey P. J., Buote D. A., Brighenti F., Flohic H. M. L. G., Gastaldello F., Mathews W. G., 2012, ApJ, 748, 11
\bibitem[\protect\citeauthoryear{Kormendy \& Bender}{2009}]{Kormendy09} Kormendy J., Bender R., 2009, ApJ, 691L, 142
\bibitem[\protect\citeauthoryear{Krajnovi\'c et al.}{2006}]{Krajnovic06} Krajnovi\'c D., Cappellari M., de Zeeuw P. T., Copin Y., 2006, MNRAS, 366, 787
\bibitem[\protect\citeauthoryear{Kroupa}{2001}]{Kroupa11} Kroupa P., 2001, MNRAS, 322, 231
\bibitem[\protect\citeauthoryear{Lagan\'a et al.}{2013}]{Lagana13} Lagan\'a T. F., Martinet N., Durret F., Lima Neto G. B., Maughan B., Zhang Y.-Y., 2013, A\&A, 555, 66L
\bibitem[\protect\citeauthoryear{Law et al.}{2015}]{Law15} Law D.R., et al., 2015, AJ, 150, 19
\bibitem[\protect\citeauthoryear{Law et al.}{2016}]{Law16} Law D. R., et al., 2016, AJ, 152, 83
\bibitem[\protect\citeauthoryear{Lelli et al.}{2017}]{Lelli17} Lelli F., McGaugh S. S., Schombert J. M., Pawlowski M. S., 2017, ApJ, 836, 152
\bibitem[\protect\citeauthoryear{Li et al.}{2016}]{Li16} Li H., Li R., Mao S., Xu D., Long R. J., Emsellem E., 2016, MNRAS, 455, 3680 
\bibitem[\protect\citeauthoryear{Li et al.}{2017a}]{Li17a} Li H., Ge J., Mao S., et al., 2017a, ApJ, 838, 77L
\bibitem[\protect\citeauthoryear{Li et al.}{2017b}]{Li17b} Li H., Mao S., Emsellem E., Xu D., Springel V., Krajnovi\'c D., 2017b, arXiv: 1709.03345
\bibitem[\protect\citeauthoryear{McGaugh et al.}{2016}]{McGaugh16} McGaugh S. S., Lelli F., Schombert J. M., 2016, Phys. Rev. Lett., 117, 201101
\bibitem[\protect\citeauthoryear{Penoyre et al.}{2017}]{Penoyre17} Penoyre Z., Moster B. P., Sijacki D., Genel S., 2017, MNRAS, 468, 3883
\bibitem[\protect\citeauthoryear{Poci, Cappellari, \& McDermid}{2017}]{Poci17} Poci A., Cappellari M., McDermid R.~M., 2017, MNRAS, 467, 1397 
\bibitem[\protect\citeauthoryear{Salpeter}{1955}]{Salpeter55} Salpeter E. E., 1955, ApJ, 121, 161
\bibitem[\protect\citeauthoryear{S\'anchez-B\'azquez}{2006}]{Sanchez06} S\'anchez-Bl\'azquez P., et al., 2006, MNRAS, 371, 703
\bibitem[\protect\citeauthoryear{Su et al.}{2013}]{Su13} Su Y., White R. E., Miller E. D., 2013, ApJ, 775, 89
\bibitem[\protect\citeauthoryear{Thomas et al.}{2011}]{Thomas11} Thomas J., et al., MNRAS, 415, 545
\bibitem[\protect\citeauthoryear{van Dokkum et al.}{2017}]{vanDokkum17}	van Dokkum P., Conroy C., Villaume A., Brodie J., Romanowsky A. J., 2017, ApJ, 841, 68
\bibitem[\protect\citeauthoryear{Vazdekis et al.}{2010}]{Vazdekis10} Vazdekis A., S\'anchez-Bl\'azquez P., Falc\'on-Barroso J., Cenarro A. J., Beasley M. A., Cardiel N., Gorgas J., Peletier R. F., 2010, MNRAS, 404, 1639
\bibitem[\protect\citeauthoryear{Weijmans et al.}{2014}]{Weijmans14} Weijmans A.-M., et al., 2014, MNRAS, 444, 3340
\bibitem[\protect\citeauthoryear{Yan et al.}{2016a}]{Yan16a} Yan R., et al., 2016a, AJ, 151, 8                 
\bibitem[\protect\citeauthoryear{Yan et al.}{2016b}]{Yan16b} Yan R., et al., 2016b, AJ, 152, 197

\end{thebibliography}

\end{document}